\documentclass[aps,prc,twocolumn, superscriptaddress,showkeys,nofootinbib]{revtex4-1}  

\usepackage[utf8]{inputenc}

\usepackage{graphicx,color}
\usepackage{amsmath,amssymb,amsfonts}
\usepackage{hyperref}

\usepackage{xspace}	

\newcommand{\roots} {\mbox{$\sqrt{\textit{s}_{NN}}$}\xspace}

\def  \vn         {\mbox{$\textit{v}_{n}$}\xspace} 
\def  \first       {\mbox{$\textit{v}_{1}$}\xspace}
\def  \second  {\mbox{$\textit{v}_{2}$}\xspace}
\def  \third      {\mbox{$\textit{v}_{3}$}\xspace}
\def  \fourth    {\mbox{$\textit{v}_{4}$}\xspace}

\def  \higher    {\mbox{$\textit{v}_{n > 3}$  }\xspace}
\def  \etas       {\mbox{$\eta / \textit{s}$  }\xspace}

\def \sc23  {\mbox{$\mathrm{SC}(2,3)$   }\xspace}
\def \sc24  {\mbox{$\mathrm{SC}(2,4)$   }\xspace}

\def \nsc23 {\mbox{$\mathrm{NSC}(2,3)$}\xspace}
\def \nsc24 {\mbox{$\mathrm{NSC}(2,4)$}\xspace}

\begin{document}
\title{Investigations of the  linear and non-linear flow harmonics \\ using the A Multi-Phase Transport model}
\medskip
\author{Niseem~Magdy} 
\email{niseemm@gmail.com}
\affiliation{Department of Physics, University of Illinois at Chicago, Chicago, Illinois 60607, USA}

\begin{abstract}
%
%
The  Multi-Phase Transport model (AMPT) is used to study the effects of the parton-scattering cross-sections ($\sigma_{pp}$) and hadronic re-scattering on the linear contributions to the flow harmonic \fourth, the non-linear response coefficients, and the correlations between different order flow symmetry planes in Au+Au collisions at 200~GeV. The model results, which agree with current experimental measurements, indicate that the higher-order flow harmonics are sensitive to the $\sigma_{pp}$ variations. However, the non-linear response coefficients and the correlations between different order flow symmetry planes are $\sigma_{pp}$ independent. These results suggest that further detailed experimental measurements which span a broad range of collision systems and beam energies could serve as an additional constraint for the theoretical models' calculations.

%
\end{abstract}
\maketitle
\section{Introduction}
Numerous experimental investigations of heavy-ion collisions at the Large Hadron Collider (LHC) and the Relativistic Heavy Ion Collider (RHIC) indicate the creation of the matter predicted by Quantum Chromodynamics (QCD), called Quark-Gluon Plasma (QGP)~\cite{Shuryak:1978ij,Shuryak:1980tp,Muller:2012zq}, in these collisions. 
Many of the previous and current experimental investigations in heavy-ion collisions are aimed at a better understanding of the QGP  transport properties (especially, \etas)~\cite{Shuryak:2003xe,Romatschke:2007mq,Luzum:2008cw,Bozek:2009dw,Acharya:2019vdf,Acharya:2020taj,Adam:2020ymj}.

A wealth of knowledge about \etas was obtained by studying the anisotropic flow observables.
The anisotropic flow measurements are expected to display the viscous hydrodynamic response to the initial spatial distribution formed in the collision's early stages~\cite{Heinz:2001xi,Hirano:2005xf,Huovinen:2001cy,Hirano:2002ds,Romatschke:2007mq,Luzum:2011mm,Song:2010mg,Qian:2016fpi,Magdy:2017ohf,Magdy:2017kji,Schenke:2011tv,Teaney:2012ke,Gardim:2012yp,Lacey:2013eia}.  
Anisotropic flow can be represented via the Fourier expansion~\cite{Poskanzer:1998yz} of the distribution of the  particle azimuthal angle $\phi$,   as;
\begin{eqnarray}
\label{eq:1-1}
\frac{dN}{d\phi} \propto 1+2\sum^{\infty}_{n=1}\textit{v}_{n} cos\left( n \left(\phi - \psi_{RP}\right)   \right)  
\end{eqnarray}
where \vn represents the value of the $n^{th}$ order flow harmonic, and $\psi_{RP}$ is the reaction plane defined by the beam direction and impact parameter~\cite{Poskanzer:1998yz}. 
The \first is called directed flow, \second is named elliptic flow, and \third is termed triangular flow, etc.  The earlier investigations of flow correlations and fluctuations~\cite{STAR:2018fpo,ALICE:2016kpq,Adamczyk:2017hdl,Qiu:2011iv, Adare:2011tg, Aad:2014fla, Aad:2015lwa,Magdy:2018itt,Alver:2008zza,Alver:2010rt, Ollitrault:2009ie} and higher-order flow coefficients \higher~\cite{Magdy:2019ojv,Adam:2019woz,Magdy:2018itt,Adamczyk:2017ird,Magdy:2017kji,Adamczyk:2017hdl,Alver:2010gr, Chatrchyan:2013kba}, have guided us toward a better understanding of the QGP properties.

In a hydrodynamic-like scenario, anisotropic flow is driven by the spatial anisotropy of the initial-state energy density,  characterized by the complex eccentricity vector $\mathcal{E}_{n}$~\cite{Alver:2010dn,Petersen:2010cw,Lacey:2010hw,Teaney:2010vd,Qiu:2011iv}:
\begin{eqnarray}
\mathcal{E}_{n} & \equiv &  \varepsilon_{n} e^{i {\textit{n}} \Phi_{n} }  \\  \nonumber
&\equiv  &
  - \frac{\int dx\,dy\,\textit{r}^{n}\,e^{i {\textit{n}} \varphi}\, \textit{E}(r,\varphi)}
           {\int dx\,dy\,\textit{r}^{n}\,\textit{E}(r,\varphi)}, ~(\textit{n} ~>~ 1),
\label{epsdef1}
\end{eqnarray}
where  $\varepsilon_{n}$ and $\mathrm{\Phi_{n}}$ are the value and azimuthal direction of the n$^{th}-order$ eccentricity vector, $x=r~\cos\varphi$, $y=r~\sin\varphi$, $\varphi$ is the spatial azimuthal angle, and ${\textit{E}}(r,\varphi)$ is the initial anisotropic energy density profile~\cite{Teaney:2010vd,Bhalerao:2014xra,Yan:2015jma}. 
To a good degree the  \second~\cite{Adamczyk:2016gfs,Adamczyk:2015fum,Magdy:2018itt,Adamczyk:2015obl,Wei:2018xpm} and \third~\cite{Adamczyk:2016exq,Adam:2019woz} are linearly related to $\varepsilon_{{{2}}}$ and $\varepsilon_{{{3}}}$, respectively~\cite{Song:2010mg, Niemi:2012aj,Gardim:2014tya, Fu:2015wba,Holopainen:2010gz,Qin:2010pf,Qiu:2011iv,Gale:2012rq,Liu:2018hjh}: 
\begin{eqnarray}
\label{eq:1-2}
v_{n} = \kappa_{n} \varepsilon_{n},
\end{eqnarray}
where $\kappa_{n}$ is expected to be sensitive to the QGP \etas~\cite{Adam:2019woz,Heinz:2013th}. 
Although the lower- and higher-order flow harmonics are emerging from linear response to the same-order eccentricity, the higher-order harmonics, \higher, in addition, contains a non-linear response to the lower-order eccentricities $\varepsilon_{{{n = 2, 3}}}$~\cite{Teaney:2012ke,Bhalerao:2014xra,Yan:2015jma,Gardim:2011xv}:
\begin{eqnarray}
V_{4}  &=&  v_{4} e^{i 4 \psi_{4}} = \kappa_{4} \varepsilon_{4} e^{4i\Phi_{4}} + \kappa_{4}^{'} \varepsilon^{2}_{2} e^{4i\Phi_{2}},\nonumber \\
           &=&  V_{4}^{\rm Linear} +  V_{4}^{\rm Non Linear}, \nonumber \\
           &=&  V_{4}^{\rm Linear}  +  \chi_{4,22} V_{2} V_{2},  \label{eq:1-3} 
\end{eqnarray}
where $\kappa_{4}^{'}$ carry information about the medium properties as well as the coupling between the lower and higher order eccentricity harmonics\cite{Liu:2018hjh}. The terms $\textit{V}^{Linear}_{4}$ and $\textit{V}^{\rm Non Linear}_{4}$ are the linear and the non-linear contributions respectively. The $\chi_{4,22}$ represents the non-linear response coefficients. 

The non-linear contribution to $\textit{V}_{4}$ will display the correlation between different order flow symmetry planes. This correlation is expected to shed light on the heavy-ion collisions' initial stage dynamics~\cite{Bilandzic:2013kga, Bhalerao:2014xra, Aad:2015lwa, ALICE:2016kpq, STAR:2018fpo,Zhou:2016eiz, Qiu:2012uy,Teaney:2013dta, Niemi:2015qia, Zhou:2015eya,Zhao:2017yhj}. The linear and non-linear contributions to higher-order flow harmonics have been discussed in several experimental publications~\cite{Magdy:2020bij,CMS:2019nct,ALICE:2019xkq,ALICE:2020sup,ALICE:2021adw,Adam:2020ymj,ATLAS:2015qwl}.  In addition, they are also widely discussed in several phenomenological studies using different transport and hydrodynamic models~\cite{Magdy:2020bhd,Yang:2019sye,Liu:2018hjh,Bozek:2017thv,Chattopadhyay:2017bjs,Zhao:2017yhj}.

 In this work, I investigated the influence of the parton-scattering
cross-sections ($\sigma_{pp}$) and the hadronic re-scattering on the linear and non-linear contributions to the $\textit{V}_{4}$ as well as the coupling constant ($\chi_{4,22}$) and the correlations between different order flow symmetry planes ($\rho_{4,22}$).
 Here, an important objective is to gain improved insights on the sensitivity of these measures to the initial-state geometry and the medium's properties. Note that in transport models~\cite{Lin:2004en} $\sigma_{pp}$ have been related to the final state transport confection \etas~\cite{Xu:2011fi,Nasim:2016rfv,Solanki:2012ne}.
%
%
The current sensitivity study, conducted within the AMPT model~\cite{Lin:2004en} framework, could lend important insights on ongoing beam energy measurements~\cite{Magdy:2020bij} of the linear and non-linear contributions to $v_{n>3}$ designed to constrain the dependence of $\eta/s$ on $T$ and  $\mu_B)$.

The paper is organized as follows. Section~\ref{sec:2} describes the theoretical model and the analysis method employed to compute  the linear and non-linear contribution of $\textit{V}_{4}$. Sec.~\ref{sec:3}, reports the results from this analysis, and is followed by a summary presented in Sec.~\ref{sec:4}.

\section{Methodology}
\label{sec:2}

\subsection{The AMPT model}
\label{sec:2a}
This investigation is conducted with simulated events for Au--Au collisions at $\sqrt{s_{\rm NN}}$ = 200~GeV, obtained using the AMPT~\cite{Lin:2004en} model. Calculations were made  for charged hadrons in the transverse momentum  span $0.2 < p_T < 2.0$ GeV/$c$ and the pseudorapidity acceptance $|\eta|$ $<$ $1.0$. The latter selection mimics the acceptance of the STAR experiment at RHIC. 

The AMPT  model~\cite{Lin:2004en} is widely employed to investigate the physics of the relativistic heavy-ion collisions at LHC and RHIC energies~\cite{Lin:2004en,Ma:2016fve,Haque:2019vgi,Bhaduri:2010wi,Nasim:2010hw,Xu:2010du,Magdy:2020bhd,Guo:2019joy}. 
In this study, simulations were performed with the string melting option in the AMPT model both on and off. In a string melting scenario hadrons produced using the HIJING model are converted to their valence quarks and anti-quarks, and their evolution in time and space is then evaluated by the ZPC parton cascade model~\cite{Zhang:1997ej}.
The essential components of AMPT are (i) an HIJING model~\cite{Wang:1991hta,Gyulassy:1994ew} initial parton-production stage, (ii) a parton-scattering stage, (iii) hadronization through coalescence  followed by (iv)  a hadronic interaction stage~\cite{Li:1995pra}.  
In the parton-scattering stage the employed parton-scattering cross-sections are estimated according to;
\begin{eqnarray} \label{eq:21}
\sigma_{pp} &=& \dfrac{9 \pi \alpha^{2}_{s}}{2 \mu^{2}},
\end{eqnarray}
where $\alpha_{s}$ is the QCD coupling constant and $\mu$ is the screening mass in the partonic matter. 
They generally give the expansion dynamics of the A--A collision systems~\cite{Zhang:1997ej}; 
Within the AMPT framework, the initial value of \etas (evaluated at the beginning of heavy-ion collision) can be related to $\sigma_{pp}$ see Refs~\cite{Xu:2011fi,Nasim:2016rfv,Solanki:2012ne}.

%
%
%
%


In this work, Au+Au collisions at $\sqrt{s_{\rm NN}}=$ 200~GeV, were simulated with AMPT version ampt-v2.26t9b for a fixed value of $\alpha_{s}$ = 0.47, but varying $\mu$ in the range 2.26 -- 4.2~$fm^{-1}$~\cite{Xu:2011fi,Nasim:2016rfv}. 
The presented AMPT sets in this work are given in Tab.~\ref{tab:1}.
%
\begin{table}[h!]
\begin{center}
 \begin{tabular}{|c|c|c|c|}
 \hline 
 AMPT-set      &         $\sigma_{pp}$      &     String Melting  Mechanism       \\
  \hline
  Set-1        &           6.1              &                     OFF                        \\
 \hline
  Set-2        &           6.1              &                     ON                         \\
 \hline 
  Set-3        &           2.7              &                     ON                         \\
 \hline 
  Set-4        &           1.8              &                     ON                         \\
 \hline
\end{tabular} 
\caption{The summary of the AMPT sets used in this work.}
\label{tab:1}
\end{center}
\end{table}

The results presented in sec.~\ref{sec:2b}, were obtained for minimum bias Au--Au collisions at $\sqrt{s_{\rm NN}}=$ 200 GeV.  A total of approximately 4.0, 5.0, 4.0, and 3.0~M events of Au--Au collisions were generated with AMPT Set-1, Set-2, Set-3, and Set-4, respectively.
\subsection{Analysis Method}
 \label{sec:2b}

The two- and multi-particle cumulants methods~\cite{Bilandzic:2010jr,Bilandzic:2013kga,Gajdosova:2017fsc,Jia:2017hbm}, are used in this work.  
The two- and multi-particle cumulants can be constructed in terms of n$^{th}$ flow vectors ($Q_{n}$) magnitude.
 The  $Q_{n}$ are given as:
{\footnotesize 
\begin{eqnarray}\label{eq:21-1}
Q_{n,k}              &=&  \sum^{M}_{i=1} \omega^{k}_{i} e^{in\varphi_{i}},
\end{eqnarray}
}
where $M$ is the total number of particles in an event and  $\omega_{i}$  is the $\textit{i}^{th}$ particle weight. We also introduce the sum over the particles weight as:
{\footnotesize 
\begin{eqnarray}\label{eq:21-2}
S_{p,k}              &=&  \left[  \sum^{M}_{i=1} \omega^{k}_{i}  \right]^{p}.
\end{eqnarray}
}

Using Eqs.(\ref{eq:21-1}, \ref{eq:21-2}) the two-, three-,  and four-particle correlations were constructed using the two-subevents cumulant methods~\cite{Jia:2017hbm},  with  $|\Delta\eta| = |\eta_{a} - \eta_{b}|  > 0.7$ ($\eta_{a}~ > 0.35$ and $\eta_{b}~ < -0.35$).
{\small
\begin{eqnarray}\label{eq:21-3}
\left(  v^{\rm Inclusive}_{n}  \right)^{2}     &=&   \sum^{N_{ev}}_{i=1} (\mathcal{M}_{2})_{i} \left\langle  2_{n} \right\rangle_{i} / \sum^{N_{ev}}_{i=1} (\mathcal{M}_{2})_{i}, \nonumber \\  
\left\langle  2_{n} \right\rangle      &=&  \dfrac{Q^{\eta_{a}}_{n,1} \left(   Q^{\eta_{b}}_{n,1} \right)^{*} }{\mathcal{M}_{2}} \nonumber \\
\mathcal{M}_{2}&=& S^{\eta_{a}}_{1,1}  S^{\eta_{b}}_{1,1},
\end{eqnarray}
\begin{eqnarray}\label{eq:21-4}
C_{n+m,nm}     &=&   \sum^{N_{ev}}_{i=1} (\mathcal{M}_{3})_{i} \left\langle  3_{n,m} \right\rangle_{i} / \sum^{N_{ev}}_{i=1} (\mathcal{M}_{3})_{i}, \nonumber\\
\left\langle  3_{n,m} \right\rangle      &=&  \dfrac{ \left(    Q^{\eta_{a}}_{n,1}  Q^{\eta_{a}}_{m,1}  -   Q^{\eta_{a}}_{n+m,2}      \right)          \left(    Q^{\eta_{b}}_{n+m,1}   \right)^{*}    }
{ \mathcal{M}_{3} }, \nonumber \\
\mathcal{M}_{3}&=& \left(  S^{\eta_{a}}_{2,1}   - S^{\eta_{a}}_{1,2}    \right)   S^{\eta_{b}}_{1,1},
\end{eqnarray} 
\begin{eqnarray}\label{eq:21-5}
\langle v_{n}^{2} v_{n}^{2}  \rangle      &=&   \sum^{N_{ev}}_{i=1} (\mathcal{M}_{4})_{i} \left\langle  4_{n} \right\rangle_{i} / \sum^{N_{ev}}_{i=1} (\mathcal{M}_{4})_{i}, \nonumber\\
\left\langle  4_{n} \right\rangle      &=&  \dfrac{ \left(    Q^{\eta_{a}}_{n,1}  Q^{\eta_{a}}_{n,1}  -  S^{\eta_{a}}_{1,2} Q^{\eta_{a}}_{2n,1}      \right)          \left(    Q^{\eta_{b}}_{n,1}  Q^{\eta_{b}}_{n,1}  -  S^{\eta_{b}}_{1,2} Q^{\eta_{b}}_{2n,1}      \right)^{*}    }
{ \mathcal{M}_{4} }, \nonumber \\
\mathcal{M}_{4}&=& \left(  S^{\eta_{a}}_{2,1}   - S^{\eta_{a}}_{1,2}    \right)   \left(  S^{\eta_{b}}_{2,1}   - S^{\eta_{b}}_{1,2}    \right).
\end{eqnarray} 
}
The benefit of using the two-subevents technique is that it assists the reduction of the nonflow correlations due to resonance decays, Bose-Einstein correlations, and the fragments of individual jets~\cite{Magdy:2020bhd}.


The  non-linear contribution to \fourth can be given as~\cite{Yan:2015jma,Bhalerao:2013ina}:
\begin{eqnarray}\label{eq:2-4}
v_{4}^{\rm Non Linear} &=&  \frac{C_{4,22}} {\sqrt{\langle \mathrm{v_2^2 v_2^2 }\rangle}}, 
\end{eqnarray}
%
%
and the linear contribution to \fourth can be expressed as:
\begin{eqnarray}\label{eq:2-5}
v_{4}^{Linear}  &=& \sqrt{ (v^{\rm Inclusive}_{4})^{\,2} - (v^{\rm Non Linear}_{4})^{\,2}  }.
\end{eqnarray}
Equation (\ref{eq:2-5}) suggests that $\textit{v}_{4}^{\rm Non Linear}$ and $\textit{v}_{4}^{\rm Linear}$ are independent~\cite{Yan:2015jma,Magdy:2020bhd}.

The non-linear response coefficient ($\chi_{4,22}$), 
which quantify the  mode-coupling contributions to the \fourth, is defined as:
\begin{eqnarray}\label{eq:2-7}
\chi_{4,22} &=& \frac{v^{\rm Non Linear}_{4}} {\sqrt{\langle  v_{2}^{2} \, v_{2}^{2} \rangle}}.
\end{eqnarray}

The correlations between different order flow symmetry planes ($\rho_{4,22}$)~\cite{Acharya:2017zfg}  can be given as:
\begin{eqnarray}\label{eq:2-6}
\rho_{4,22} &=& \frac{v^{\rm Non Linear}_{4}}{v^{\rm Inclusive}_{4}}  = \langle  \cos (4 \Psi_{4} - 2 \Psi_{2} - 2 \Psi_{2}) \rangle.
\end{eqnarray}
\section{Results and discussion}\label{sec:3}
\begin{figure}[!h] 
\includegraphics[width=1.0 \linewidth, angle=-0,keepaspectratio=true,clip=true]{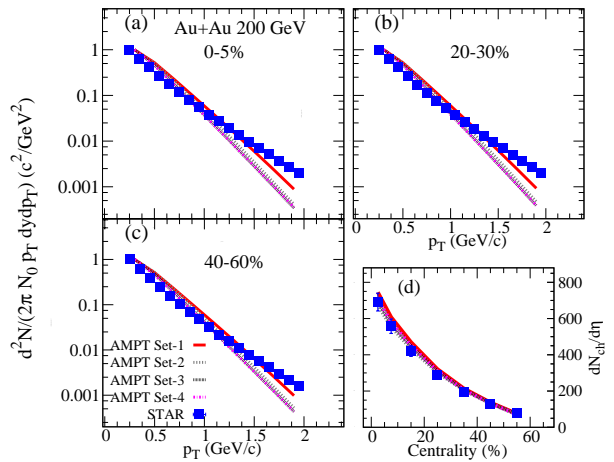}
\vskip -0.4cm
\caption{The charged-particle transverse momentum spectra and multiplicity density of Au+Au collision at 200 GeV.  The solid points represent the experimental data reported by the STAR collaboration~\cite{STAR:2003fka,STAR:2008med}.
  }\label{fig:1x}
\end{figure}

Before I present the AMPT model flow harmonics results, I first compare the AMPT model transverse momentum spectra and the charged-particle multiplicity density with the experimental measurements~\cite{STAR:2003fka,STAR:2008med}.
 Fig. ~\ref{fig:1x} shows the transverse momentum spectra (a)-(c) and the charged-particle multiplicity density (d) for Au+Au collisions at 200~GeV from the AMPT model.
 The model results are also compared to the STAR collaboration measurements (solid points)~\cite{STAR:2003fka,STAR:2008med}.  They indicate that the AMPT results agree with the STAR measurements at low $p_T$ ($<$1.3~GeV/$c$) but fails to explain the data for larger $p_T$.
 By contrast, AMPT shows good overall agreement with the experimental charged-particle multiplicity density.

%
\begin{figure}[!h] 
\includegraphics[width=1.0 \linewidth, angle=-0,keepaspectratio=true,clip=true]{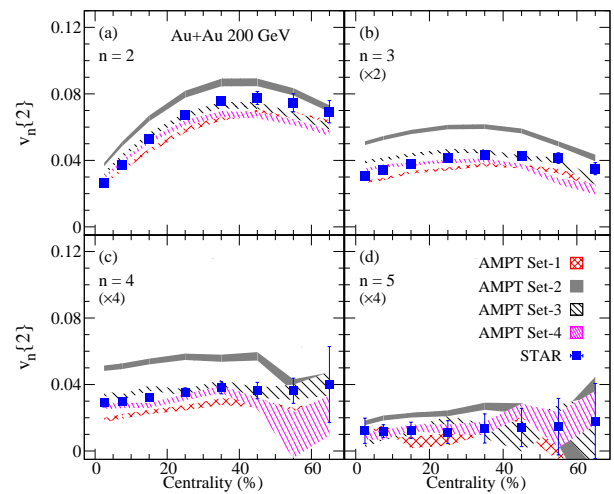}
\vskip -0.4cm
\caption{Comparison of the experimental and simulated centrality-dependent $v_{n}\lbrace 2\rbrace$ for Au+Au collisions at 200~GeV. The solid points represent the experimental data reported by the STAR collaboration~\cite{Adamczyk:2016exq,Adamczyk:2017hdl}.
  }\label{fig:1}
\end{figure}
%
\begin{figure}[hbt]
    \includegraphics[width=1.0 \linewidth, angle=-0,keepaspectratio=true,clip=true]{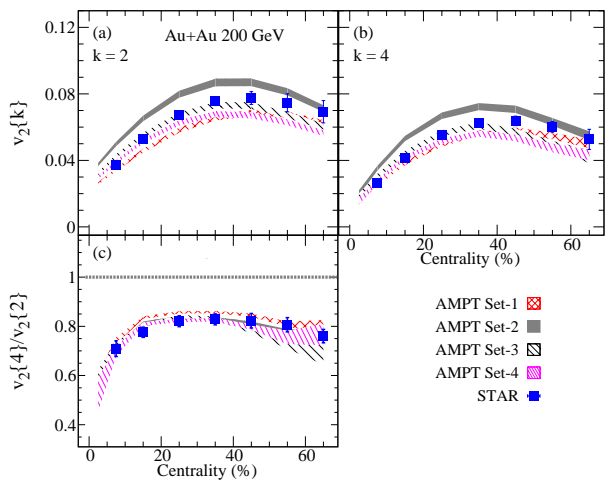}
      \vskip -0.4cm
    \caption{    
    Centrality dependence of $v_{2}\lbrace 2\rbrace$ (a), $v_{2}\lbrace 4\rbrace$ (b) and $v_{2}\lbrace 4\rbrace / v_{2}\lbrace 2\rbrace$ (c) computed with the AMPT model for Au--Au collisions at 200~GeV. The solid points represent the experimental data reported by the STAR collaboration~\cite{Adam:2019woz,Adamczyk:2017hdl}.
  }  \label{fig:2}
\end{figure}
\begin{figure}[hbt]
  \vskip -0.2cm
    \centering
\includegraphics[width=1.0 \linewidth, angle=-0,keepaspectratio=true,clip=true]{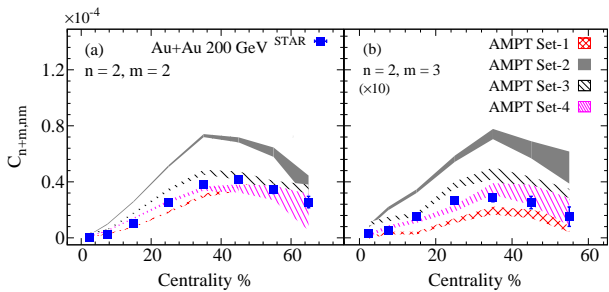} 
\vskip -0.4cm
     \caption{
		Comparison of the three-particle correlators, $C_{4,22}$ and $C_{5,23}$ centrality dependence, from the AMPT model for Au--Au collisions at 200~GeV. The solid points represent the experimental data reported by the STAR collaboration~\cite{Adam:2020ymj}.
               }
    \label{fig:3} 
  \vskip -0.2cm
\end{figure}
\begin{figure*}[ht] 
    \includegraphics[width=0.99 \linewidth, angle=-0,keepaspectratio=true,clip=true]{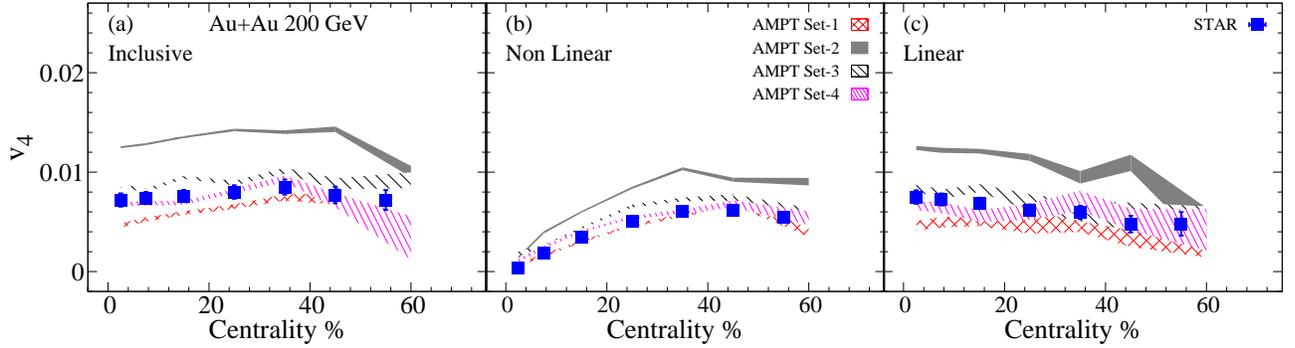}
    \vskip -0.4cm
    \caption{
    Centrality dependence of  the inclusive, non-linear and linear \fourth  obtained with the two-subevents cumulant method from the AMPT model for Au--Au collisions at 200~GeV. The solid points represent the experimental data reported by the STAR collaboration~\cite{Adam:2020ymj}.
  } \label{fig:4}
    \vskip -0.0cm
\end{figure*}
\begin{figure}[hbt]
  \vskip -0.3cm
\includegraphics[width=0.99 \linewidth, angle=-0,keepaspectratio=true,clip=true]{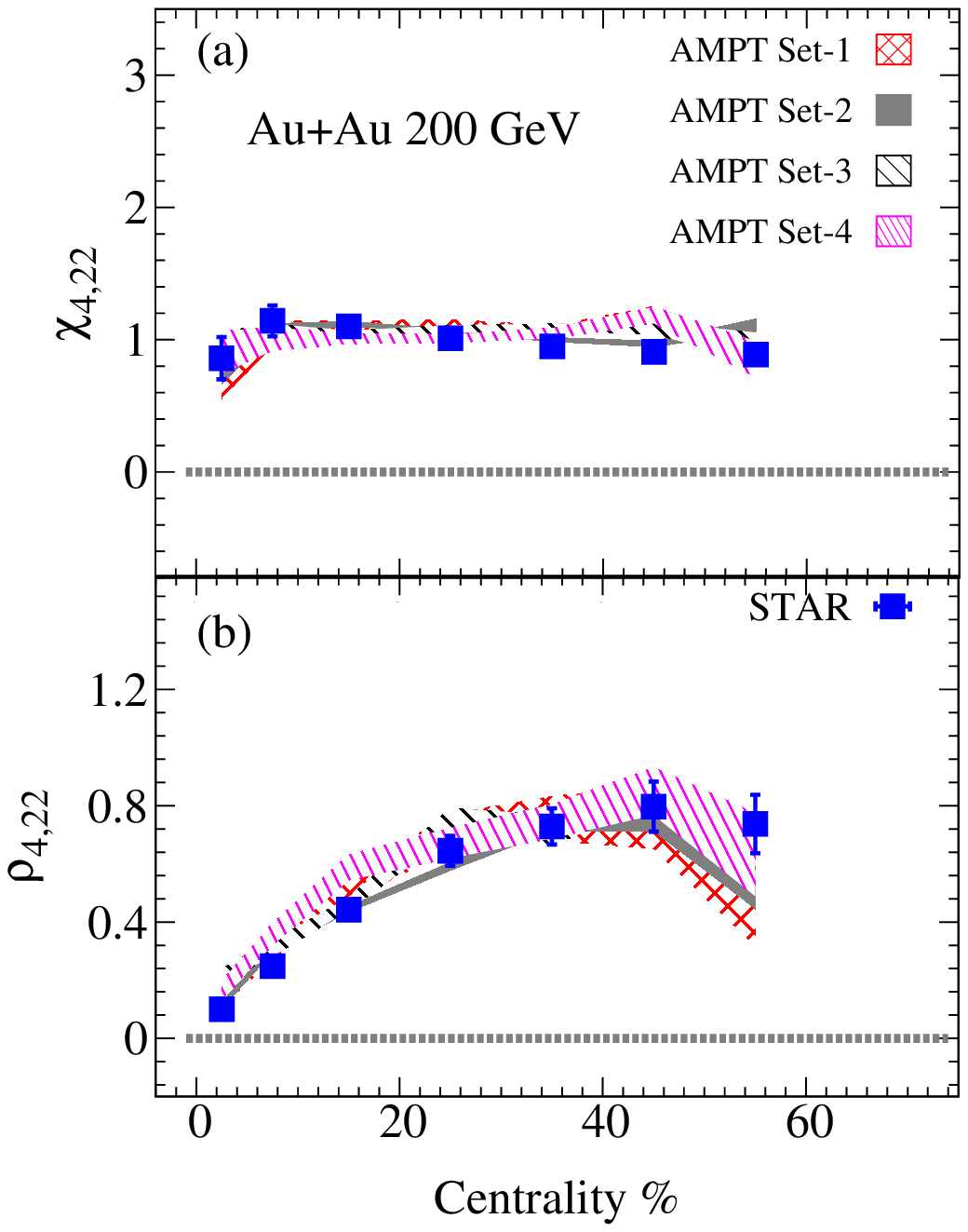}
  \vskip -0.5cm
 \caption{
Comparison of the $\chi_{4,22}$ and $\rho_{4,22}$ obtained from the AMPT model for Au--Au collisions at 200~GeV, as a function of centrality. The solid points represent the experimental data reported by the STAR collaboration~\cite{Adam:2020ymj}.~\label{fig:5}
}
 \vskip -0.2cm
\end{figure}

The extraction of the linear and the non-linear contributions to \fourth relies on the two- and multiparticle correlations. Therefore, it is instructive to investigate the dependence of these variables on the model parameters tabulated in Table~\ref{tab:1}.
Fig. ~\ref{fig:1} shows a comparison of the centrality dependence of  $v_{n}\lbrace 2 \rbrace$ (a)-(d)  for Au--Au collisions at 200~GeV from the AMPT model.  The presented $v_{n}\lbrace 2 \rbrace$ from the AMPT exhibit a sensitivity to both $\sigma_{pp}$ and whether or not string melting is turned on. They also indicate similar patterns to the data reported by the STAR collaboration~\cite{Adamczyk:2016exq,Adamczyk:2017hdl} (solid points) depending on the model parameters given in Table~\ref{tab:1}.

Figure~\ref{fig:2} shows a comparison of the centrality dependence of  $v_{2}\lbrace 2\rbrace$ (a), $v_{2}\lbrace 4\rbrace$ (b) and the ratios $v_{2}\lbrace 4\rbrace/v_{2}\lbrace 2\rbrace$ (c) for the AMPT model. The results presented in Fig.~\ref{fig:2} panels (a) and (b)  show that AMPT $v_{2}\lbrace k\rbrace$ (k=2 and 4) are sensitive to the model parameters tabulated in Tab.~\ref{tab:1}.They also give similar values and trends to those measured by the STAR experiment~\cite{Adams:2004bi}.
The ratios $v_{2}\{4\}/ v_{2}\{2\}$, shown in panel (c), work as a figure of merit for the elliptic flow fluctuations strength; $v_{2}\{4\}/ v_{2}\{2\} \sim 1.0$ correspond to minimal flow fluctuations and decreasing values of $v_{2}\{4\}/ v_{2}\{2\} < 1.0$ for larger flow fluctuations.
The estimated ratios, which are in good agreement with the experimental ratios, are to within $\sim $2\% insensitive to the model setups given in Tab.~\ref{tab:1}, implying that the flow fluctuations in the AMPT model are eccentricity-driven and are roughly a constant fraction of $v_{2}\{2\}$. 

The centrality dependence of the three-particle correlators, $C_{4,22}$ and $C_{5,23}$,  are shown in Fig. ~\ref{fig:3} for Au+Au collisions at 200~GeV from the AMPT model. 
These results indicate that $C_{4,22}$ and $C_{5,23}$ are strongly sensitive to the $\sigma_{pp}$ and also sensitive to whether or not string melting is turned on. They also show patterns similar to the experimental data reported by the STAR collaboration~\cite{Adam:2020ymj}.

The centrality dependence of the inclusive, linear and non-linear \fourth for Au+Au collisions at \roots = 200~GeV from the AMPT model are shown in Fig.~\ref{fig:4}. The presented measurements indicate that the linear contribution of \fourth which is the dominant contribution to the inclusive \fourth in central collisions has a weak centrality dependence.
The presented results show that the inclusive, linear and non-linear \fourth are sensitive to the $\sigma_{pp}$. These results are compared to STAR collaboration measurements~\cite{Adam:2020ymj}. The AMPT model simulations indicate similar qualitative patterns to the \fourth  measured by the STAR collaboration~\cite{Adam:2020ymj}.


The centrality dependence of the non-linear response coefficients, $\chi_{4,22}$, for Au+Au collisions at \roots = 200~GeV, from the AMPT model is presented in Fig.~\ref{fig:5}(a).
The $\chi_{4,22}$ results indicate a weak centrality and $\sigma_{pp}$ dependence,  which implies that (i) the  non-linear \fourth centrality dependence arises from the lower-order flow harmonics and (ii)  the $\chi_{4,22}$ is dominated by initial-state eccentricity couplings.
The  $\chi_{4,22}$ from the AMPT model shows a good agreement with the $\chi_{4,22}$ measured by the STAR experiment~\cite{Adam:2020ymj}.

Figure~\ref{fig:5}(b) shows the centrality dependence of the correlations of the event plane angles, $\rho_{4,22}$, in Au--Au collisions at \roots = 200~GeV from the AMPT model.
The  $\rho_{4,22}$ results imply stronger event plane correlations in peripheral collisions for all presented AMPT sets. However, the calculated magnitudes of $\rho_{4,22}$ are shown and found to be independent of $\sigma_{pp}$. Such observation suggests that the correlation of event plane angles are dominated by initial-state correlations.
The  $\rho_{4,22}$ from the AMPT model indicate a good agreement with the $\rho_{4,22}$ measured by the STAR collaboration~\cite{Adam:2020ymj}.


\section{Conclusions} \label{sec:4}
In summary, I have presented extensive AMPT model studies to evaluate the $\sigma_{pp}$ dependence of the linear and non-linear contributions to the \fourth, non-linear response coefficients, and the correlations of the event plane angle for Au--Au collisions at \roots = 200~GeV. 
The presented AMPT calculations indicate a strong centrality dependence for the non-linear \fourth in contrast, the linear \fourth shows a soft centrality dependence.
In addition, \fourth calculations show a strong sensitivity to the $\sigma_{pp}$ and whether or not string melting is turned on. 
The dimensionless parameters $\chi_{4,22}$ and $\rho_{4,22}$ show magnitudes and trends which are $\sigma_{pp}$ independent, suggesting that the correlations of event plane angles, as well as the non-linear response coefficients, are dominated by initial-state effects. 
Based on these AMPT model calculations, I conclude that conducting further detailed experimental measurements of the linear and non-linear response to the higher-order flow harmonics over a broad range of beam-energies and for different collision systems could serve as an additional constraint for theoretical models.

%
\section*{Acknowledgments}
This research is supported by the US Department of Energy, Office of Nuclear Physics (DOE NP),  under contracts DE-FG02-94ER40865.
%
%

\bibliography{ref} 
\end{document}